\documentclass[%
twocolumn,
prd,
showpacs,preprintnumbers,
nofootinbib,
amsmath,amssymb,
]{revtex4}

\usepackage{graphicx}
\usepackage{dcolumn}
\usepackage{bm}
\usepackage{hyperref}
\usepackage{color}
\usepackage{breakurl}

\newcommand{\fatc}{{\bf c}}
\newcommand{\fatmu}{{\boldsymbol \mu}}
\newcommand{\TeV}{\,\text{TeV}}
\newcommand{\kpc}{\,\text{kpc}}
\newcommand{\GeV}{\,\text{GeV}}
\newcommand{\sr}{\,\text{sr}}
\newcommand{\s}{\,\text{s}}
\newcommand{\cm}{\,\text{cm}}

\newcommand{\noise}{\mathcal{N}}
\newcommand{\signal}{\mathcal{S}}
\newcommand{\bg}{\mathcal{B}}
\newcommand{\fex}{\textit{e.g.}~}
\newcommand{\ie}{\textit{i.e.}~}

\begin{document}

\preprint{DESY 11-097}
\preprint{MPP-2011-60}

\title{On the Relevance of Sharp Gamma-Ray Features for Indirect Dark Matter Searches}

\author{Torsten Bringmann}
\email{torsten.bringmann@desy.de}
\author{Francesca Calore}
\email{francesca.calore@desy.de}
\affiliation{{II.} Institute for Theoretical Physics, University of Hamburg,
Luruper Chaussee 149, DE-22761 Hamburg, Germany}
\author{Gilles Vertongen}
\email{gilles.vertongen@desy.de}
\affiliation{Deutsches Elektronen-Synchrotron (DESY), Notkestrasse 85, 22603 Hamburg, Germany}
\author{Christoph Weniger}
\email{weniger@mppmu.mpg.de}
\affiliation{Max-Planck-Institut f\"ur Physik, F\"ohringer Ring 6, 80805 Munich, Germany}

\date{13 Oktober 2011}

\begin{abstract}
  Gamma rays from the annihilation of dark matter particles in the Galactic
  halo provide a particularly promising means of indirectly detecting dark
  matter. Here, we demonstrate that pronounced spectral features at energies
  near the dark matter  particles' mass, which are a generic prediction for
  most models, can significantly improve the sensitivity of gamma-ray
  telescopes to dark matter  signals. We derive projected limits on such
  features  (including the traditionally looked-for line signals) and show
  that they can be much more efficient in constraining the nature of dark
  matter than the model-independent broad spectral features expected at lower
  energies.
\end{abstract}


\maketitle


\section{Introduction}
Indirect dark matter (DM) searches  aim at seeing an excess in cosmic rays
from the annihilation or decay of DM in the Galactic
halo~\cite{Bertone:2004pz}.  Gamma rays play a pronounced role in this respect
because they are produced rather copiously and directly trace their sources as
they propagate essentially unperturbed through the galaxy.  Powerful currently
operating telescopes like Fermi LAT~\cite{Atwood:2009ez},
H.E.S.S.~\cite{Aharonian:2006pe}, MAGIC \cite{Albert:2007xh} or VERITAS
\cite{veritas} now start to constrain viable DM models and next generation
instruments like the planned CTA \cite{Consortium:2010bc} will be able to dig
quite a bit into the underlying parameter space of particle physics models, in
a way complementary to both direct DM detection and searches at the CERN LHC
\cite{Bergstrom:2010gh}.

Very often, indirect searches focus on \emph{secondary photons} from the
fragmentation of annihilation products, mostly via
$\pi^0\rightarrow\gamma\gamma$. The resulting spectrum is rather
model-independent and would manifest itself as a  broad bump-like excess over
the expected background at energies considerably lower than the DM mass
$m_\chi$.  Convincingly claiming a DM detection based on the observation of
such a feature-less signal will generically  be difficult.

In many  models, however, \emph{pronounced spectral features} are expected at
the kinematic endpoint $E_\gamma=m_\chi$
include monochromatic gamma-ray lines \cite{Bergstrom:1997fj}, sharp steps or
cut-offs \cite{Birkedal:2005ep,Bergstrom:2004cy} as well as pronounced bumps
\cite{Bringmann:2007nk}. The type and strength of these features are
intricately linked to the particle nature of DM; a detection would thus not
only allow a convincing discrimination from astrophysical backgrounds but also
to determine important DM model parameters (in particular, but not necessarily
limited to, the value of $m_\chi$).  So far, only line-signals have explicitly
been searched for~\cite{line_searches}---despite the fact that they are
loop-suppressed and thus generically subdominant compared to other spectral
signatures \cite{Bringmann:2007nk}.

Here, we present a general method to search for such features and show that
these, indeed, help significantly to discriminate DM signals from
astrophysical backgrounds. This allows  us to derive very competitive
(projected) limits on both  the annihilation rate and nature of DM,  which we
believe will be very useful for DM searches.


\section{Method}
\label{sec:method}
The defining aspect of the above-mentioned spectral features in the DM-induced
gamma-ray emission is an abrupt change of the flux as function of energy; in
the extreme cases of gamma-ray lines or cut-offs, \textit{e.g.}, the
corresponding energy range would simply be given by the energy resolution
$\Delta E/E$ of the instrument.  The basic idea that we will adopt here,
following traditional gamma-ray line searches~\cite{line_searches}, is
therefore to concentrate the search for spectral features on a small sliding
energy window $[E_0, E_1]$, with $E_0\!<\! m_\chi\! <\! E_1$ and
$\varepsilon\equiv E_1/E_0\sim\mathcal{O}(1$--$10)$.  An important advantage
of using small values for $\varepsilon$ is that gamma-ray fluxes with
astrophysical origin can often be very well described by a simple power-law.
In that case, a corresponding fit to the data allows an effective
determination of the background at the statistical limit,  greatly reducing
uncertainties related to astrophysical sources. 

For deriving constraints on spectral features within the sliding energy
window, we will use a binned profile likelihood method~\cite{Rolke:2004mj}.
To this end, we split  $[E_0,E_1]$  in many energy bins $\Delta E_i$ and
define a likelihood function  $L(\fatmu|{\fatc}) = \Pi_{i} P_{\mu_i}(c_i)$,
where $\mu_i$  ($c_i$) denotes the expected (observed) count number in  bin
$i$ and $P_\mu$ is the Poisson probability distribution with mean $\mu$.  Introducing
the background normalization $\beta$, its spectral slope $\gamma$ and the
normalization of the DM signal $\alpha$, we have
\begin{equation}
  \frac{\mu_i}{t_{\rm obs}} =\!
  \int_{\Delta E_i}\!\!\!\!\!\!\!\!dE\int\!\! dE'\, \mathcal{D}_{E, E'}
  A_\text{eff}(E')\!\left[\alpha\frac{dN_\chi}{dE'} + \beta
  E'^{-\gamma}\right],\!
  \label{eqn:mu}
\end{equation}
where $t_{\rm obs}$ is the time of observation, $A_{\rm eff}$ the effective
area and $\mathcal{D}_{E, E'}$  the energy dispersion of the instrument (in
the following taken to be Gaussian). Maximizing $L(\fatmu|{\fatc})$ for a
given data set $\fatc$  results in  best-fit values of the model parameters
$\alpha$, $\beta$ and $\gamma$ within the considered window $[E_0,E_1]$.
\emph{Upper limits} on the signal strength at the $95.5\%$ C.L.~can then be
derived by increasing $\alpha$ from its best-fit value until $-2\log L$
(maximized with respect to $\beta$ and $\gamma$) has changed by 4. On the
other hand, a \emph{detection} at the $5\sigma$ level (neglecting trial
factors) could be claimed if the best-fit $-2\log L$ values for
background-only and background-plus-signal fits differ by at least 25.

Our main assumption here is that the astrophysical background \emph{locally}
takes the form of a power law. Obviously, this approximation can break down in
case of  large window sizes $\varepsilon$, depending on the collected
statistics and, to first order, on the intrinsic curvature of the background
flux $\kappa\equiv d^2\log (dJ_\text{BG}/dE)/(d\log E)^2$: a change in the
spectral index by $\Delta\gamma$, e.g., implies roughly
$|\kappa|\sim\Delta\gamma^2/4$ at the transition point;  $\kappa$ could,
however, also be affected by systematic uncertainties in $A_\text{eff}$. We
will derive constraints on the maximally allowed window size
$\varepsilon_\text{max}$ by requiring  that these effects do not significantly
alter the resulting DM limits.

\section{Choice of target and instrument specifications}

\begin{table}[t]
  \begin{ruledtabular}
    \begin{tabular}{lrrrr}
        & $A_\text{eff}(1\TeV)$  & $\Delta E/E(1\TeV)$ & $\epsilon_p$ & $t_\text{obs}$\\
      \colrule\\[-3mm]
      IACT1 &  0.18\ km$^2$ & 15\% & $10^{-1}$ & 50\,h\\
      IACT2 &  2.3\ km$^2$ & 9\% & $10^{-2}$ & 100\,h\\
      IACT3 &  23\ km$^2$ & 7\% & $10^{-3}$ & 5000\,h\\
    \end{tabular}
  \end{ruledtabular}
  \caption{IACT benchmark models that, from top to bottom,
  roughly correspond to the H.E.S.S.~\cite{Aharonian:2006pe}, the future
  CTA~\cite{Consortium:2010bc} and the proposed DMA~\cite{Bergstrom:2010gh}
  telescope characteristics.}
  \label{tab:BM}
\end{table}

For concreteness, we will in the following focus on  observations of the
Galactic center  region with Imaging Atmospheric Cherenkov Telescopes (IACTs).
We consider the benchmark scenarios  summarized in Tab.~\ref{tab:BM}, which
roughly correspond to the telescope characteristics of the currently operating
H.E.S.S.~\cite{Aharonian:2006pe}, the future CTA~\cite{Consortium:2010bc}
and---as the most optimistic choice for indirect DM searches---the proposed
Dark Matter Array (DMA)~\cite{Bergstrom:2010gh}.  We implement the energy
dependence of the effective area $A_\text{eff}$ as given in
Ref.~\cite{Aharonian:2006pe} (Ref.~\cite{InternalNote}) for H.E.S.S.~(CTA) and
take $A_\text{eff}^{\rm DMA}=10\cdot A_\text{eff}^{\rm CTA}$.  The proton,
gamma-ray and electron efficiencies $\epsilon_{p,\gamma,e^-}$ in all three
scenarios as well as the energy resolution $\Delta E/E$ in case of
H.E.S.S.~and DMA are taken to be energy independent; for CTA we adopt results
from Ref.~\cite{Consortium:2010bc}. We will use
$\epsilon_\gamma=\epsilon_{e^-}=0.8$ throughout and assume that the proposed
DMA can reject protons with efficiencies $\epsilon_p\approx 10^{-3}$.  

For the background, we take into account cosmic-ray fluxes of electrons
\cite{electrons} and protons \cite{Hoerandel:2002yg, Fegan:1997db}, the
diffuse gamma-ray flux \cite{Aharonian:2006au} and the source HESS
J1745-290~\cite{Aharonian:2006wh} at (or very close to) the Galactic center. A
summary and more detailed description can be found in the Appendix; there, we
also discuss which choice of target region $\Delta\Omega$  optimizes the
signal-to-noise ratio $\signal/\noise$ (see also Ref.~\cite{Serpico:2008ga}).
For the Einasto and NFW DM profiles, with parameters as in
Ref.~\cite{profiles}, we will adopt a relatively small region
$\Delta\Omega=2^\circ\times2^\circ$ around the Galactic center; larger regions
would weaken the signal-to-background ratio, $\signal/\bg$. In case of a
strong point source-like enhancement of the DM signal from the Galactic center
(\textit{e.g.}~through the effect of the super-massive black
hole~\cite{Gondolo:1999ef,Berezinsky:1992mx} or adiabatic
compression~\cite{Blumenthal:1985qy, Gnedin:2003rj, Gustafsson:2006gr}), it is
favorable to focus on even much smaller target regions. As an example, we
consider the case of an adiabatically compressed (AC) profile
\cite{Gnedin:2003rj,Gustafsson:2006gr} for which we choose a target region of
$\Delta\Omega=0.2^\circ\times0.2^\circ$.

\begin{figure}[t]
  \includegraphics[width=\columnwidth]{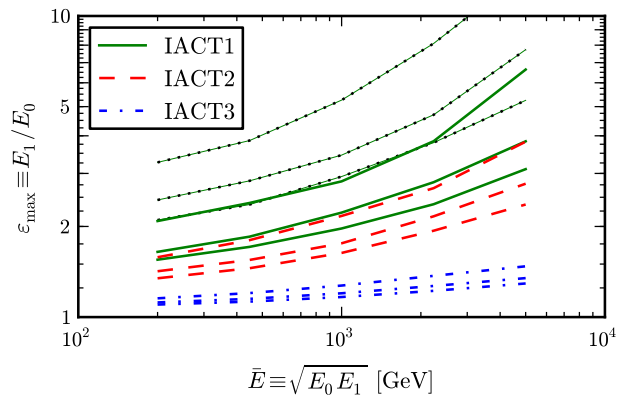}
  \caption{Maximal sliding energy window size $\varepsilon_\text{max}$ as
  function of the  window position $\bar E$. We show for different intrinsic
  background curvatures $\kappa_\text{max}=0.1, 0.2, 0.3$ (top to bottom) the
  window sizes above which DM limits are affected by more than $50\%$. The
  \textit{dotted lines} show for IACT1 $\varepsilon_\text{max}$ for which a
  power-law ansatz would still give a good fit to the  background.}
  \label{fig:windowsize} 
\end{figure}

Let us now derive values for the  tolerable window size
$\varepsilon_\text{max}$ in presence of maximal background curvatures
$\kappa_\text{max}$. Using mock data sets with  $\kappa=\pm\kappa_\text{max}$,
we compare average DM limits (on the  DM models introduced below) obtained
when using the power-law ansatz for the background with the limits obtained
when incorporating a (fixed) curvature $\kappa$ in the background fit. In
Fig.~\ref{fig:windowsize} we display as function of the sliding energy window
position $\bar{E}\equiv\sqrt{E_0 E_1}$, and for different curvatures
$\kappa_\text{max}$, the values of $\varepsilon_\text{max}$ above which the DM
limits are affected by more than $50\%$.  For comparison, the dotted lines
show for IACT1 the values for $\varepsilon_\text{max}$ below which the
power-law fit still appears to be in good agreement with the curved background
(using as criterion that for at least $80\%$ of the mock data sets the
$p$-value of the power-law fit is larger than $0.05$): obviously, a good
quality of the power-law fit alone does not automatically exclude sizeable
effects on the DM limits. Therefore, $\emph{a priori}$ assumptions on
$\kappa_\text{max}$ are indispensable; in our case, we employ
$|\kappa|\leq\kappa_\text{max}\approx0.2$ -- which we checked to be satisfied
for the background  we adopt here --  to determine optimal logarithmic window
sizes for IACT1 and IACT2 according to Fig.~\ref{fig:windowsize} (for IACT3,
see below).


\section{Dark matter spectral signatures}

\begin{table}[t]
  \begin{ruledtabular}
    \begin{tabular}{lrrrrr}
        & DM particle  & $m_\chi^{\rm th}$ & $\langle\sigma v\rangle^{\rm th}$ & relevant & spectral\\
        && [TeV] & [cm$^3$s$^{-1}$] & channel & feature\\
      \colrule\\[-3mm]
      $\gamma\gamma$  &  any WIMP & $\mathcal{O}$(0.1--10) & $\mathcal{O}(10^{-30})$& $\gamma\gamma$ & line\\
      KK  & $B^{(1)}$  & 1.3 & $1\cdot10^{-26}$ & $\ell^+\ell^-\gamma$ & FSR step\\
      BM3 &  neutralino & 0.23 & $9\cdot10^{-29}$ & $\ell^+\ell^-\gamma$ & IB bump\\
      BM4 &  neutralino & 1.9 & $3\cdot10^{-27}$ & $W^+W^-\gamma$ & IB bump\\
    \end{tabular}
  \end{ruledtabular}
  \caption{DM benchmark models used in our analysis as examples for the typical 
   spectral endpoint features to be expected in WIMP annihilations. For these particular models, we also state the annihilation channel that is most important in this context, as well as  mass and total annihilation rate for thermally produced DM. See text for further details about the DM models and Fig.~\ref{fig:DMspectra} for the corresponding photon spectra.}
  \label{tab:DM}
\end{table}

The DM signal flux  from a sky region $\Delta\Omega$ is given by
\begin{equation}
 \frac{dJ_\chi}{dE} \equiv
\alpha\frac{dN_\chi}{dE} = \frac{\langle\sigma v\rangle}{8\pi
  m_\chi^2}\int_{\Delta \Omega} \! \! \! \!d\Omega \int_\text{l.o.s.}  \! \!\! \! \! ds\
  \rho_\chi(r(s, \Omega))^2\ \frac{dN_\chi}{dE}\;,
  \label{eqn:flux}
\end{equation}
where $\langle\sigma v\rangle$ is the annihilation rate, ${dN_\chi}/{dE}$  the
differential number of photons per annihilation,  $\rho_\chi(r)$  the Galactic
DM profile and $s$ runs over the line-of-sight.  For any  photon spectrum, 
 and a given value of the dark matter particles' mass, 
we can now derive limits on $\alpha$ (aka $\langle\sigma v\rangle$) by scanning 
over all possible values of $m_\chi$ and applying
 the method described in detail in Section \ref{sec:method}. 
As a technical remark, we found that the best limits are actually obtained by choosing  the center of the
sliding energy window  to lie slightly off-set from the
kinematic endpoint $E=m_\chi$ of DM spectra at or slightly below which we
expect to see the features we are looking for. For the  instrument specifications and 
background model that we adopted here, in particular, the optimal choice turned out to be $\bar E=\varepsilon^{-0.25}m_\chi$ 
(not for line signals, however, for which we take $\bar E=m_\chi$).

In the following, we will discuss three types of typical endpoint features
that arise from radiative corrections to the tree-level annihilation process.
The most striking spectral signature, in terms of a possible discrimination
from a power-law background, is a \emph{gamma-ray line} at $E_\gamma = m_\chi$
($E_\gamma = m_\chi[1 - m_{Z/H}^{2}/4 m_\chi^{2}]$), which would result from
the direct annihilation of DM into $\gamma \gamma$ ($Z \gamma$ or $H\gamma$)
\cite{Bergstrom:1997fj}. Generically, for thermal cross sections of DM in the form of weakly interacting massive particles (WIMPs), the
annihilation rate is expected to be of the order of  $\langle\sigma
v\rangle_{\rm line}\sim\alpha_{\rm em}^2\times\langle\sigma v\rangle_{\rm
tree}\sim10^{-30}{\rm cm}^3{\rm s}^{-1}$, but in some cases much stronger line
signals are possible \cite{hisanoSF,IDHM, Mambrini}. 

As an example for a  step-like feature we use the gamma-ray spectrum
\cite{Bergstrom:2004cy} expected from annihilating \emph{Kaluza-Klein (KK) DM}
in models of universal extra dimensions \cite{Servant:2002aq}. In the minimal
version of these models, the DM particle is the $B^{(1)}$, {\ie}the first KK
excitation of the weak hypercharge gauge boson, and the correct relic density
is obtained for $m_{B^{(1)}}\sim1.3\,$TeV \cite{Belanger:2010yx}.  Its total
gamma-ray annihilation spectrum $dN/dx$ (with $x\equiv E/m_\chi$) at high
energies is dominated by final state radiation (FSR) off lepton final states and
turns out to be essentially independent of $m_{B^{(1)}}$ and other model
parameters.

Pronounced bump-like features at $E \simeq m_\chi$ may arise from internal
bremsstrahlung (IB) in the annihilation of \emph{neutralino DM}
\cite{Bringmann:2007nk}.  While these spectra are in general highly
model-dependent, we follow here a simplified approach by defining two spectral
templates $dN/dx$ (which we take to be independent of $m_\chi$) by referring
to  neutralino benchmark models introduced in Ref.~\cite{Bringmann:2007nk}.
Here, BM3 is a typical example for a neutralino in the stau co-annihilation
region,  where photon emission from virtual sleptons greatly enhances $dN/dx$;
BM4 refers to a situation in which IB from $W^\pm$ final states  dominates. We
note that the Sommerfeld effect could strongly enhance these features, in
particular in the case of BM4, in the same way as pointed out in
Ref.~\cite{hisanoSF} for line signals.

\begin{figure}[t!]
  \includegraphics[width=\columnwidth]{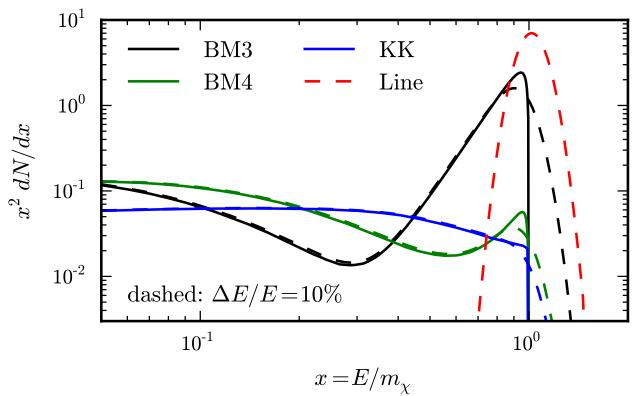}
  \caption{Photon spectra for the DM benchmark models of Tab.~\ref{tab:DM}. Dashed lines 
  show the same spectra, smeared with a Gaussian of width $\Delta x/x=0.1$ to give a rough indication of how well a detector with such an energy resolution would in principle be able to discriminate 
   these models from  astrophysical (power-law) backgrounds, as well as from each other.}
  \label{fig:DMspectra} 
\end{figure}

\begin{figure*}[t]
  \includegraphics[width=\linewidth]{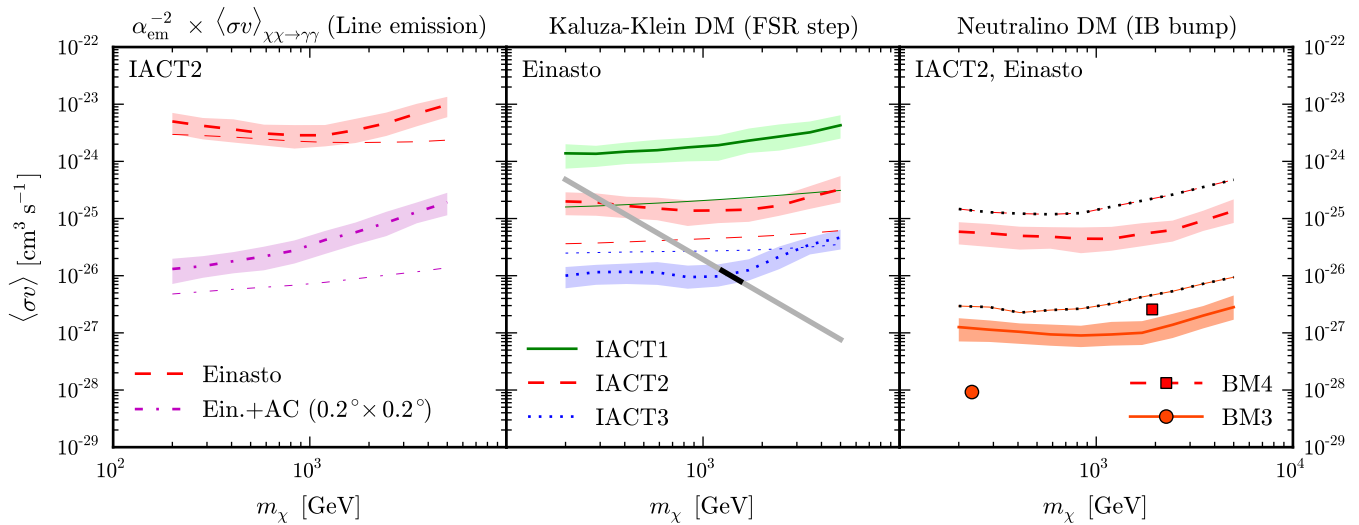}
  \caption{\textit{Thick lines:} Expected $2\sigma$ upper limits on
  $\langle\sigma v\rangle$ for  selected DM models, DM profiles and
  observational scenarios; bands indicate the variance of these limits.
  \textit{Thin lines:} Spectral feature of DM signal has
  $\signal/\bg\approx1\%$ (after convolution with energy dispersion).  The
  \textit{left panel} shows limits on gamma-ray lines, rescaled by a
  loop-factor of $\alpha_\text{em}^{-2}$ for better comparison.  In the
  \textit{central panel}, the gray band indicates the expected $\langle\sigma
  v\rangle$ for KK DM, the black part being compatible with the observed relic
  density. In the \textit{right panel}, we indicate the adopted neutralino
  benchmark points, and the \textit{dotted lines} show the projected $5\sigma$
  sensitivity.}
  \label{fig:CS} 
\end{figure*}

In Tab.~\ref{tab:DM}, we shortly summarize the properties of the 
DM benchmark models described above, including for completeness 
the actual DM mass and total 
annihilation rate needed to obtain the observed relic density for thermally produced DM. Note, however, that we 
essentially treat these values as free parameters in our analysis and that we are rather 
interested in the spectral shape of the annihilation signal, represented by $dN/dx$; in 
Fig.~\ref{fig:DMspectra} we show these spectra for a direct comparison.

\section{Limits and discussion} 

In Fig.~\ref{fig:CS} we show our results for the expected $2\sigma$ upper
limits (thick lines) on the above DM models as well as the variance of these
limits among the $300$ mock data sets that we created for this analysis.  We
find that in particular IB features in the spectrum (right panel) have the
potential to constrain the annihilation rate at least down to values typically
expected for thermal production, $\langle \sigma
v\rangle\sim3\cdot10^{-26}{\rm cm}^3{\rm s}^{-1}$, already for modest
assumptions about the DM distribution (we verified that NFW and Einasto
profiles give  similar results). This is very competitive compared to the best
current limits from ACTs that only rely on secondary photons
\cite{Abramowski:2011hc} --  though we would like to stress that these limits
provide rather complementary information on the DM nature and can thus usually
not easily be compared. 

For the case of not too strongly pronounced endpoint features (like line
signals in most models or the step for Kaluza-Klein DM), secondary photons
will usually be more powerful in constraining the total annihilation rate
$\langle\sigma v\rangle$; from the point of view of indirect DM searches,
however, the \emph{detection} of the kinematic cutoff will be much more
interesting than the detection of secondary photons since it allows to draw
firmer conclusions about the DM origin of the signal and even to determine
important parameters like $m_\chi$. For models with very large IB
contributions like BM3, on the other hand, we find that our method provides
even stronger \emph{limits} on $\langle\sigma v\rangle$ than what was obtained
by the HESS analysis of the Galactic center region assuming annihilation  into
$\bar b b$  \cite{Abramowski:2011hc}.

In case of an adiabatically compressed profile our limits could improve by two
orders of magnitude, as demonstrated for gamma-ray lines in the left panel;
under such conditions, one could even hope to constrain models with very small
annihilation rates like BM3 (recall that the annihilation rate for BM4 is
anyway affected by the Sommerfeld enhancement \cite{hisanoSF} and thus likely
considerably larger than what is shown in Fig.~\ref{fig:CS}).  As shown in the
central panel of Fig.~\ref{fig:CS}, the future CTA should be able to place
limits  about one order of magnitude stronger than currently possible, and the
proposed DMA could further improve these by another factor of ten.\footnote{Note that
for DMA, as can be seen from Fig.~\ref{fig:windowsize}, the statistics actually become so good that a spectrum with $\kappa\sim\mathcal{O}(0.1)$
curvature starts to deviate significantly from a power-law background
  already for rather small sliding energy windows.
In order to obtain reasonable limits,  we therefore included $\kappa\neq0$ as a free parameter in the fit to allow for energy 
 windows somewhat larger
than shown in Fig.~\ref{fig:windowsize}.}

When probing a specific DM model, the corresponding $\signal/\bg$ is a good
measure for the level on which spectral artefacts in the energy reconstruction
of the instrument must be understood. As can be inferred from
Fig.~\ref{fig:CS} (thin lines), most of our derived limits correspond to
moderate $\signal/\bg$ values of at least a few percent (except for IACT3),
which should be well in reach of current instruments.

Limits on gamma-ray lines as shown in Fig.~\ref{fig:CS} are usually derived
\emph{neglecting} any secondary gamma-ray component from DM
annihilation~\cite{line_searches}; this approximation, however,  breaks down
for very small branching ratio into lines  since part of the secondary
component will leak into the sliding energy window.  Assuming a dominant
annihilation into $b\bar{b}$ final states, we find that for branching ratios
into gamma-ray lines smaller than $\mathcal{O}(10^{-4})$, the presence of the
secondary flux begins to alter the derived gamma-ray line limits
significantly.  This renders a naive application of standard line-search
results on DM models with generic $\mathcal{O}(\alpha^2_{\rm em})$ branching
ratios into gamma-ray lines questionable~\cite{us}.

The dotted lines in the right panel of Fig.~\ref{fig:CS} show the projected
\textit{sensitivity} to see a $5\sigma$ signal in the IACT2 scenario
(neglecting systematics and trial factors). Such an observation should of
course be cross-checked by the non-observation of the same signature in
control regions without large DM induced fluxes.  A more detailed analysis for
detectional prospects is beyond the scope of the present work and left for a
subsequent publication~\cite{us}.

\section{Conclusions}
Gamma rays from  DM annihilation often exhibit pronounced spectral features
near photon energies close to the DM particles' mass.  Here, we have shown
that methods from gamma-ray line searches, which greatly reduce the
uncertainties related to astrophysical background fluxes, can successfully be
extended to look for such spectral features; this provides a probe of the DM
nature that is complementary to DM searches relying only on the rather
model-independent spectrum from secondary photons.

While these kind of features may generically be considered even more relevant
for the \emph{detection} of DM signals, because they would provide rather
unambiguous evidence for the DM nature of the signal as well as allow to
determine important parameters like the DM mass, we have demonstrated here
that including the spectral information may even significantly improve
\emph{limits} on DM signals; steps or bump-like IB features can, in fact, be
much more important in this respect than lines.  

We stress that while we have considered constraints for IACT observations of
the Galactic center region, the presented method is much more general and can
be applied to both other targets and other instruments; we thus expect it to
be useful for a wide range of applications in indirect DM searches.  An
obvious extension of the approach presented here, finally, is to apply it to
the detection rather than exclusion of DM signals, as well as to the
\emph{discrimination} of models \cite{us}.

\acknowledgments
We thank U.~Almeida, D.~Borla Tridon and
H.~Zechlin for useful discussions, and M.~Kakizaki for confirming that the
gamma-ray spectrum of KK DM computed in Ref.~\cite{Bergstrom:2004cy} remains essentially
unaffected by changing the DM mass such as to be compatible with the  most
recent relic density calculations \cite{Belanger:2010yx}. T.B.~and
F.C.~acknowledge support from the German Research Foundation (DFG) through
Emmy Noether grant BR 3954/1-1.

\appendix

\section{Dark matter searches in the Galactic Center region}

Imaging Atmospheric Cherenkov Telescopes (IACTs) detect gamma rays by
measuring the dim Cherenkov light produced by electromagnetic
showers through the atmosphere. Very similar showers are induced by
cosmic-ray electrons, which hence constitute a practically irreducible
background.  Proton-induced hadronic showers, on the other hand, differ in
profile and energy density and can currently be rejected with efficiencies
$\epsilon_p\sim \mathcal{O}(10^{-2}$--$10^{-1})$.  Due to their large
intrinsic fluxes, charged cosmic rays typically form the major background of
IACT observations. For the flux of cosmic-ray electrons, we take 
$ {dJ_{e^-}}/{dEd\Omega}=1.17\times10^{-11}\left({E}/{\TeV}
  \right)^{-3.9}
  (\text{GeV}\cm^2\s\sr)^{-1}
  \label{eqn:Jee}
$ 
above $1\TeV$, which hardens below 1 TeV to a spectral index of $-3.0$
\cite{electrons} (with a transition between the two fluxes that we assume to
be proportional to their generalized mean with exponent $-2$).  For the proton
flux we take
$ {dJ_p}/{dEd\Omega} = 8.73 \times 10^{-9}
  \left({E}/{\text{TeV}}\right)^{-2.71}
  (\text{GeV}\cm^2\s\sr)^{-1}
$
\cite{Hoerandel:2002yg}, which we shift to lower energies by a factor of $3$
to take into account the reduced Cherenkov light output of hadronic showers,
$E^\text{recon.}_p \approx E_p^\text{true}/3$ (see \fex
Ref.~\cite{Fegan:1997db}).

\begin{figure}[t!]
  \includegraphics[width=\columnwidth]{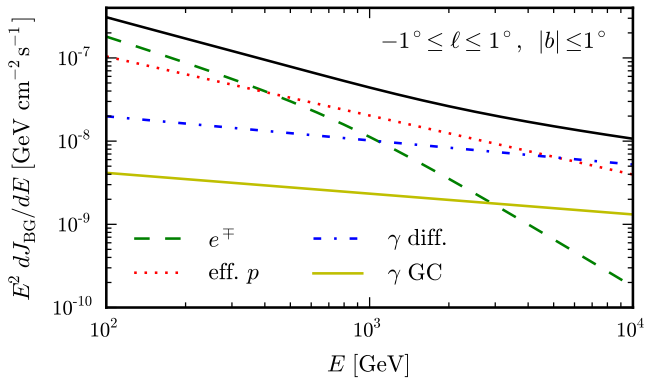}
  \caption{Summary of adopted background fluxes in a $2^\circ\times2^\circ$
  region around the Galactic center, with an effective proton flux as it enters the IACT2
  scenario. The black solid line shows the sum of all background fluxes.}
  \label{fig:fluxes} 
\end{figure}

For observations of the Galactic Center region (GC), we take as further
background into account the HESS source  J1745-290 \cite{Aharonian:2006wh},
with ${dJ_\text{HESS}}/{dE}= 2.3\times10^{-15}
\left({E}/{\text{TeV}}\right)^{-2.25} (\text{GeV}\cm^2\s)^{-1}$.  The diffuse
photon emission measured by H.E.S.S.~in a $-0.8^\circ\leq\ell\leq0.8^\circ$
and $|b|\leq0.3^\circ$ region around the GC is given by
$
  {dJ_\text{diff}}/{dE} = 5.1 \times 10^{-15}
  \left({E}/{\text{TeV}}\right)^{-2.29}
  (\text{GeV}\cm^2\s)^{-1}
$
\cite{Aharonian:2006au}. Unknown diffuse emission from outside  this region
will conservatively be accounted for by upscaling this flux by a factor of two
within our $2^\circ\times2^\circ$ target region.  We summarize all these
background contributions in Fig.~\ref{fig:fluxes}. 

The statistical significance of a spectral feature depends on the
signal-to-noise ratio $\signal/\noise$ ($\noise\simeq\sqrt{
\mathcal{B}+\signal}$) inside the target region.  The number of expected
background events $\mathcal{B}$ within a target region $\Delta \Omega$ and
energy range $\Delta E$ is calculated analogously to Eq.~\eqref{eqn:mu} by
replacing the model flux by the sum of the above background fluxes after
integrating over $\Delta \Omega$. In the same way, the number of signal events
$\signal$ follows from the DM annihilation flux as given in
Eq.~\eqref{eqn:flux}.

\begin{figure}[t]
  \includegraphics[width=\columnwidth]{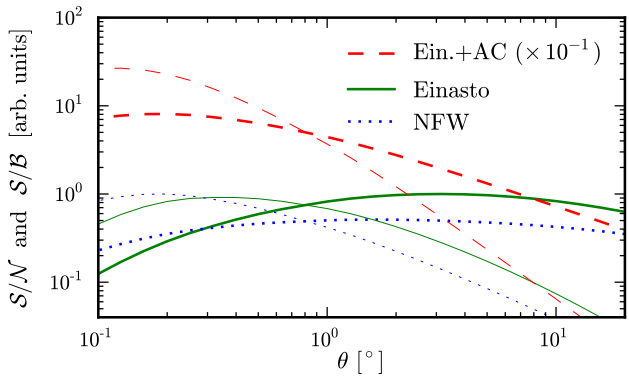}
  \caption{$\mathcal{S}/\mathcal{N}$ (\textit{thick lines}) and
  $\mathcal{S}/\mathcal{B}$ (\textit{thin lines}) of DM signal inside a
  circular region around the GC with radius $\theta$, for different DM halo
  profiles. For the Einasto profile, we also show
  the effect of adiabatic compression (AC). We assume
  $\signal\ll\bg$ and an energy threshold of $200\GeV$, but for threshold
  energies $100\GeV$--$5\TeV$ we find similar results.}
  \label{fig:sn} 
\end{figure}

In Fig.~\ref{fig:sn}, we show the $\signal/\noise$ (thick lines) of spectral
features for a circular region around the GC with radius $\theta$.  We compare
results for the standard Einasto and NFW DM profiles with parameters as in
Ref.~\cite{profiles} ({\ie}$r_s^\text{NFW}=21\kpc$, $r_s^\text{Ein.}=20\kpc$,
$\alpha=0.17$ and $\rho_\chi=0.4\GeV\cm^{-3}$ at Sun's position
$R_\odot=8.5\kpc$). As can be seen from the figure radii of a few degree are
required in order to maximize $\signal/\noise$ for these profiles (see also
Ref.\cite{Serpico:2008ga}). 

For the optimal choice of $\Delta\Omega$ one should also consider the
signal-to-background ratio $\signal/\mathcal{B}$ which is shown in
Fig.~\ref{fig:sn} for comparison (thin lines).  $\signal/\bg$ is related to
the importance of systematic instrumental effects for the statistical
analysis, {\ie}it gives an indication of how well artefacts and uncertainties
in the reconstructed energy spectrum of the instrument must be understood. In
most of our analysis, we use a relatively small
$\Delta\Omega=2^\circ\times2^\circ$ region around the GC. As can be seen in
Fig.~\ref{fig:sn}, although a larger region could improve $\signal/\noise$, it
also would imply a significantly reduced $\signal/\mathcal{B}$. 

In some cases, \textit{e.g.}~through the effect of the super-massive black
hole~\cite{Gondolo:1999ef} or adiabatic
compression~\cite{Blumenthal:1985qy,Gnedin:2003rj,Gustafsson:2006gr}, the DM
annihilation can be boosted in a region concentrated around the GC, leading to
a qualitative change in the behavior of $\signal/\noise$ with respect to the
above unboosted DM profiles. The effect of adiabatic compression is
illustrated in Fig.~\ref{fig:sn} in case of the Einasto profile (Ein.+AC),
where we exemplarily adopt the adiabatic contraction model of Gnedin
\textit{et al.}~\cite{Gnedin:2003rj} together with the best fit parameters
inferred from the hydrodynamical simulation S1 of Gustafsson \textit{et
al.}~\cite{Gustafsson:2006gr}. In this case, the profile inner slope steepens
to $\rho \sim r^{-1.12}$. For such an enhancement, DM self-annihilations start
to play a role and constrain the halo density to be at most
$\rho_{\textrm{max}} \sim m_{\chi}/\langle \sigma v \rangle
\tau_{\textrm{gal}}$ \cite{Berezinsky:1992mx}; however, the cutoff radii
obtained, $r \sim 10^{-9}\kpc$, are so small that this effect does not
influence our results. As can be seen from Fig.~\ref{fig:sn}, in the adopted
boosted scenario it is preferable to consider much smaller target regions than
in case of the above unboosted profiles; hence, we will use a
$0.2^\circ\times0.2^\circ$ region around the GC when calculating limits in
presence of adiabatic compression.

\end{document}